\documentclass[aps,prl,twocolumn,superscriptaddress]{revtex4}

\usepackage{graphicx}
\usepackage{bm}

\begin{document}

\preprint{}

\title{ High-$T_{\rm c}$ Nodeless $s_\pm$-wave Superconductivity in (Y,La)FeAsO$_{1-y}$ with $T_{\rm c}$=50 K~:\\
$^{75}$As-NMR Study}

\author{H. Mukuda}
\email[]{e-mail  address: mukuda@mp.es.osaka-u.ac.jp}
\author{S. Furukawa}
\author{H. Kinouchi}
\author{M. Yashima}
\author{Y. Kitaoka}
\affiliation{Graduate School of Engineering Science, Osaka University, Osaka 560-8531, Japan}
\author{P. M. Shirage}
\author{H. Eisaki}
\author{A. Iyo}
\affiliation{National Institute of Advanced Industrial Science and Technology (AIST), Umezono, Tsukuba 305-8568, Japan}

\date{\today}

\begin{abstract}
We report $^{75}$As-NMR study on the Fe-pnictide high-$T_{\rm c}$ superconductor Y$_{0.95}$La$_{0.05}$FeAsO$_{1-y}$ (Y$_{0.95}$La$_{0.05}$1111) with $T_{\rm c}$=50 K that includes no magnetic rare-earth elements. The measurement of the nuclear-spin lattice-relaxation rate $^{75}(1/T_1)$ has revealed that the nodeless bulk superconductivity takes place at $T_{\rm c}$=50 K while antiferromagnetic spin fluctuations develop moderately in the normal state. These features are consistently described by the multiple fully-gapped $s_\pm$-wave model based on the Fermi-surface (FS) nesting. Incorporating the theory based on band calculations, we propose that the reason that $T_{\rm c}$=50 K in Y$_{0.95}$La$_{0.05}$1111 is larger than $T_{\rm c}$=28 K in La1111 is that the FS multiplicity is  maximized, and hence the  FS nesting condition is better than that in La1111.
\end{abstract}


\maketitle

%%%%%%%%%%%%%%%%%%%%%%%%%%%%%%%%%  Introduction  %%%%%%%%%%%%%%%%%%%%%%%%%%%%
%\section{Introduction}

After the discovery of superconductivity (SC) in iron (Fe)-based oxypnictide LaFeAsO$_{1-x}$F$_x$ (hereafter denoted as La1111) with a SC transition temperature $T_{\rm c}$=26 K\cite{Kamihara2008}, $T_{\rm c}$ goes up over 50 K with the replacement of La for other magnetic rare-earth elements in $Ln$1111 ($Ln$=Sm,Nd etc.)~\cite{Ren1,Kito,Ren2}. 
Over the past four years, extensive studies have been reported on various Fe-based superconductors, pointing to the diversity of SC characteristics and normal-state electronic properties. It is believed that this diversity is associated with  the multiband/multiorbital nature of the degenerate Fe-$3d$ states in the Fe$Pn$($Pn$=As,P) layer.
In the lattice-parameter points of view, $T_{c}$ reaches a maximum of 55 K when a FeAs$_{4}$ block forms in a nearly regular tetrahedral structure in $Ln$1111~\cite{C.H.Lee}, where the height of pnictogen($h_{Pn}$) from the Fe plane and the $a$-axis length are $h_{Pn}\sim$1.38\AA~\cite{Mizuguchi} and $a\sim$3.9\AA~\cite{Ren2,Shirage,Miyazawa1}, respectively.  In order to address a possible mechanism for high-$T_{\rm c}$ SC in $Ln$1111, it is desired to gain further insight into why a structure of FeAs$_{4}$ tetrahedron is relevant with the diversity of SC characteristics and normal-state electronic properties. 
However, most $Ln$1111 are not extensively investigated by various measurements of angle-resolved-photoemission spectroscopy (ARPES), NMR, scanning tunneling spectroscopy (STS), and so on. This is partly  because high-quality single crystals with a sufficiently large size are not yet available.  The presence of $4f$-electrons derived magnetic fluctuations in $Ln$1111 has prevented us from characterizing SC and normal-state properties by means of NMR measurements \cite{Yamashita,Jeglic,Prando}. 

In this Letter,  we report for the first time a $^{75}$As-NMR study  on a high-$T_{\rm c}$=50 K-class of Fe-pnictide  superconductor Y$_{0.95}$La$_{0.05}$FeAsO$_{1-y}$ (Y$_{0.95}$La$_{0.05}$1111) that does not involve any magnetic rare earth ions. We demonstrate that antiferromagnetic spin fluctuations (AFSFs) in Y$_{0.95}$La$_{0.05}$1111 are more significant  due to the better Fermi-surface (FS) nesting condition than those in La1111 with $T_{\rm c}$=28 K,  and  as a result, Y$_{0.95}$La$_{0.05}$1111 realizes the multiple fully gapped high-$T_{\rm c}$ $s_\pm$-wave SC with $T_{\rm c}$=50 K.  

%%%%%%%%%%%%%%%%%%%%%%%%%%%%%%%   experimental    %%%%%%%%%%%%%%%%%%%%%%%%%%%%
%\section{Experimental}

\begin{figure}[tbp]
\begin{center}
\includegraphics[width=8cm]{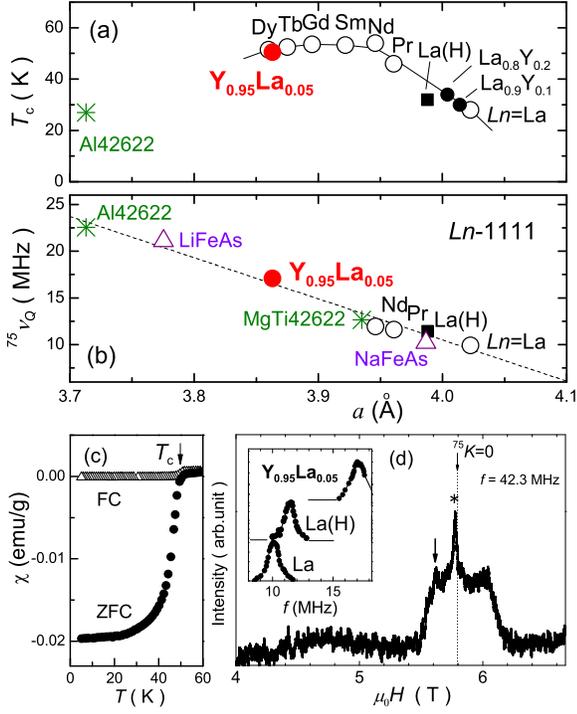}
\end{center}
\caption[]{(color online)
(a) $T_{\rm c}$s vs $a$-axis length\cite{Ren2,Shirage,Miyazawa1} and (b) $^{75}$As-NQR frequencies vs $a$-axis length for Y$_{0.95}$La$_{0.05}$1111, along with those for $Ln$1111 \cite{MukudaNQR,Yamashita_Y}, $M$42622\cite{Kinouchi,Yamamoto}, and $A$111\cite{Li,Kitagawa_Na111}. (c) SC diamagnetic susceptibility from which $T_{\rm c}$=50 K was uniquely determined. (d) $^{75}$As-NMR spectrum at $T$=50 K.  A sharp NMR spectrum denoted by $\ast$ comes from  a small amount of unreacted YAs sample, which is separately discriminated. Inset:$^{75}$As-NQR spectra of Y$_{0.95}$La$_{0.05}$1111,  La1111(H)[La(H)]($T_{\rm c}$=32 K)\cite{Yamashita_Y}, and La1111($T_{\rm c}$=28 K)\cite{MukudaNQR}. 
}
\label{Tc_a_NQR}
\end{figure}

A polycrystalline sample of Y$_{0.95}$La$_{0.05}$FeAsO$_{1-y}$ (H$_{0.15}$) was synthesized via a high-pressure synthesis technique with an addition of a small amount of hydrogen as a sort of catalyst to stabilize homogeneous samples~\cite{Shirage,Miyazawa1,Miyazawa_H}. The nominal oxygen content is $y$=$0.2\sim$0.25, but the actual oxygen content is slightly smaller than the nominal one, owing to the oxidation of starting rare-earth elements.  
The x-ray diffraction measurement indicates that the sample is composed of almost only a single phase with lattice parameters of $a$=3.863\AA\  and $c$=8.337\AA, although a tiny amount of unreacted YAs is identified. 
As shown in Fig.~\ref{Tc_a_NQR}(a), the $a$-axis length is compatible to that of $Ln$1111 with $T_{\rm c} \ge$ 50 K~\cite{Ren2,Shirage,Miyazawa1}. Bulk SC with $T_{\rm c}$=50 K for Y$_{0.95}$La$_{0.05}$1111 was determined from an onset of  SC diamagnetism in susceptibility (see Fig.~\ref{Tc_a_NQR}(c)). A coarse powder sample was used for the measurements of  nuclear spin-lattice relaxation rate $^{75}(1/T_1)$ of $^{75}$As-NMR at the field $\mu_0 H\sim 12$ T perpendicular to the $c$-axis. Note that the $^{75}$As-NMR spectrum in Y$_{0.95}$La$_{0.05}$1111 was discriminated from that of unreacted YAs. 

%----------------------- Fig.1 phase diagram ---------------------
%----------------------------------------------------------------------

%%%%%%%%%%%%%%%%%%%%%%%%%%%%% results and discussions  %%%%%%%%%%%%%%%%%%%%%%%

%%%%%%%%%%%%   NMR/NQR spectra   %%%%%%%%%%%%%%%%

Figure~\ref{Tc_a_NQR}(d) and its inset show respective $^{75}$As-NMR and NQR spectra for Y$_{0.95}$La$_{0.05}$1111. The estimated $^{75}$As-NQR frequency $^{75}\nu_{\rm Q}$ is 17.1 MHz, which is larger than $^{75}\nu_{\rm Q}$s of $Ln$1111\cite{MukudaNQR}. 
The data of $^{75}\nu_{\rm Q}$ for Y$_{x}$La$_{1-x}$1111 and $Ln$1111 are on a linear relation along with those for ($Ae_4M_2$O$_6$)Fe$_2$As$_2$ ($M$42622)\cite{Kinouchi,Yamamoto}, and $A$FeAs($A$=Li,Na) ($A$111)\cite{Li,Kitagawa_Na111}, as drawn by the dashed line in Fig.~\ref{Tc_a_NQR}(b). Note that as the $a$-axis length decreases, $^{75}\nu_{\rm Q}$ increases linearly. Since $^{75}\nu_{\rm Q}$ in proportion to an electric field gradient at As nuclear site is determined by some charge distribution around the $^{75}$As nucleus in the FeAs$_4$ tetrahedron, the increase in $h_{Pn}$ in association with the reduction in $a$-axis increases $^{75}\nu_{\rm Q}$, yielding a monotonous variation in the covalency of Fe-As bond \cite{C.H.Lee}. 

%%%%%%%%%%%%   Normal state properties   %%%%%%%%%%%%%%%%

%**************  Fig. 3****************************************
\begin{figure}[htbp]
\centering
\includegraphics[width=8cm]{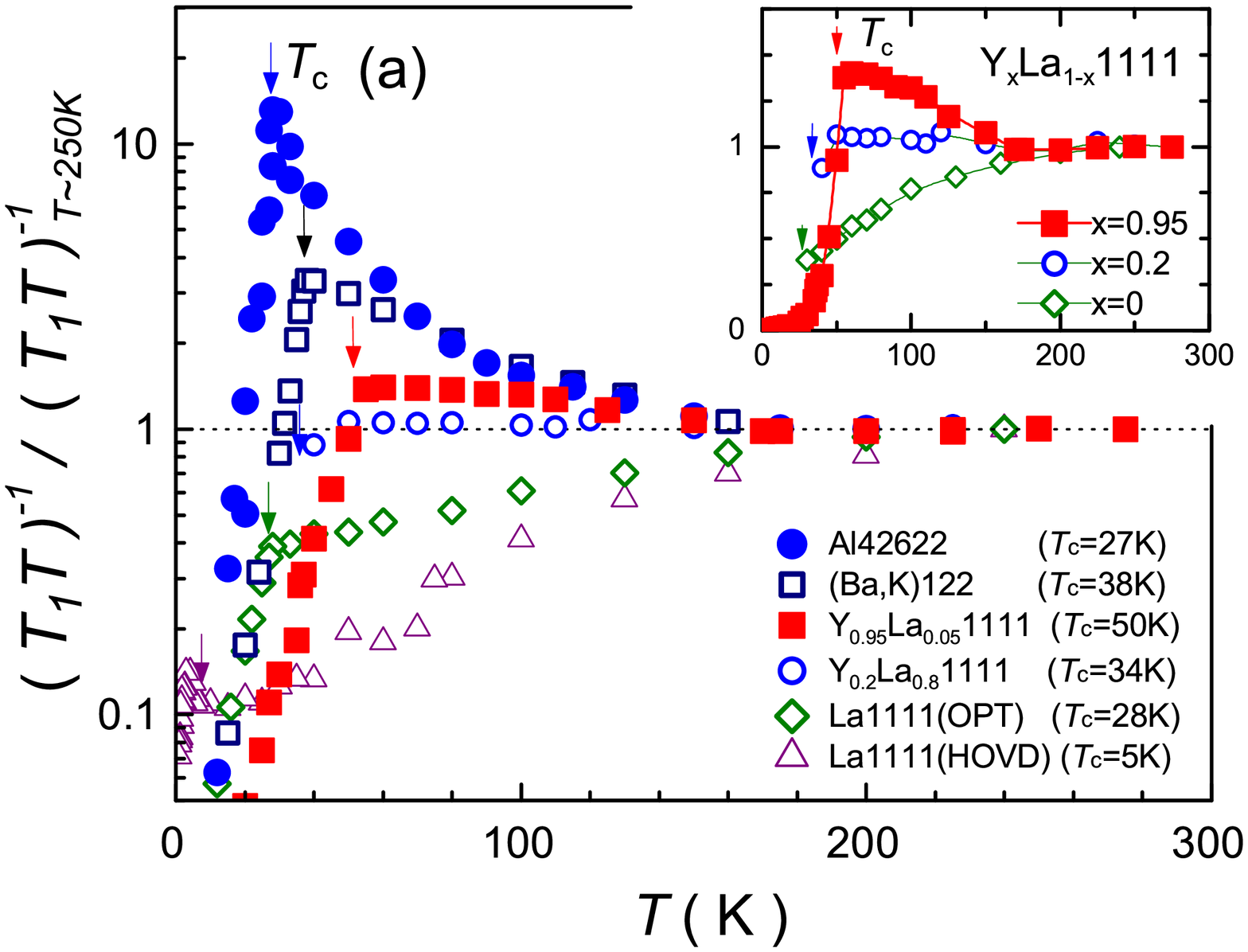}\\
\includegraphics[width=7.5cm]{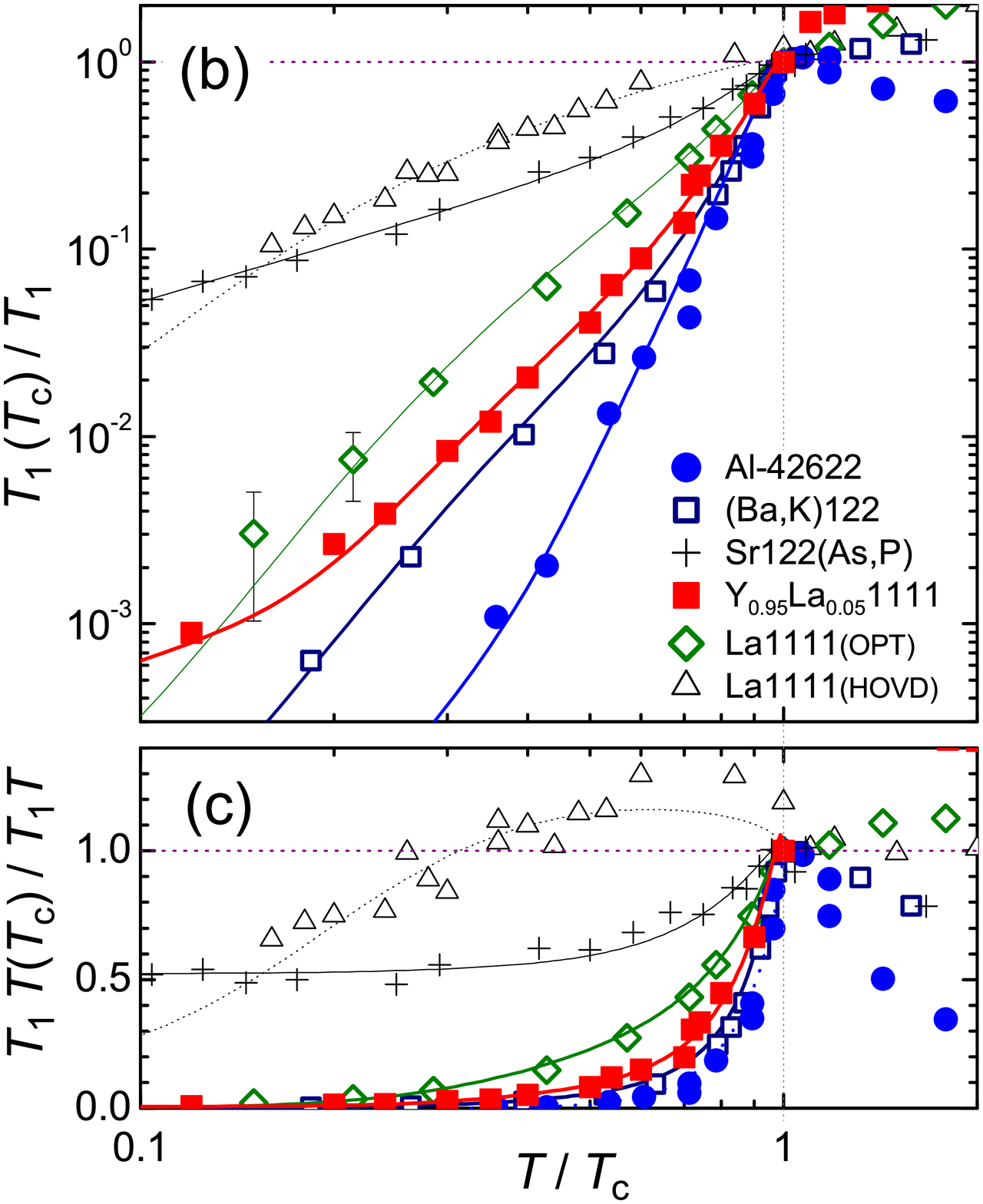}
\caption[]{(color online)
(a) $T$ dependence of  $(T_1T)^{-1}/(T_1T)^{-1}_{T\sim250 K}$ of $^{75}$As-NMR with $(T_1T)^{-1}_{T\sim250 K}$ at $T\sim$250 K, (b) $T_1(T_{\rm c})/T_1$, and (c) $T_1T(T_{\rm c})/T_1T$ as a function of $T/T_{\rm c}$ for Y$_{0.95}$La$_{0.05}$1111 (solid squares), along with the results for Y$_{0.2}$La$_{0.8}$1111 ($T_{\rm c}$=34 K)\cite{Yamashita_Y}, La1111(OPT) ($T_{\rm c}$=28 K)\cite{MukudaNQR,Yashima}, Al42622 ($T_{\rm c}$=27 K)\cite{Kinouchi}, BaK122 ($T_{\rm c}$=38 K) \cite{Yashima}, La1111(HOVD) ($T_{\rm c}$=5 K)\cite{MukudaHOVD}, and Sr122(As,P) ($T_{\rm c}$=26 K)\cite{Dulguun}.
The solid lines for figures (b) and (c) are simulations based on the multiple fully gapped $s_\pm$-wave model (see text).}
\label{fig:T1}
\end{figure}
%****************************************************************************

Figure~\ref{fig:T1}(a) shows the $T$ dependence of $^{75}1/T_1T$ (solid squares) normalized by the value at $T$=250 K for Y$_{0.95}$La$_{0.05}$1111 and other La-based compounds such as Y$_{0.2}$La$_{0.8}$1111 ($T_{\rm c}$=34 K)\cite{Yamashita_Y}, La1111(OPT) ($T_{\rm c}$=28 K)\cite{MukudaNQR,Yashima}, and La1111(HOVD) ($T_{\rm c}$=5 K)\cite{MukudaHOVD}, along with Ba$_{0.6}$K$_{0.4}$Fe$_{2}$As$_{2}$ (BaK122) ($T_{\rm c}$=38 K)\cite{Yashima} and (Ca$_4$Al$_2$O$_6$)Fe$_2$As$_2$ (Al42622) ($T_{\rm c}$=27 K)\cite{Kinouchi}.
The inset of Fig.~\ref{fig:T1}(a) presents a systematic $T$ evolution in  $^{75}1/T_1T$ for a series of  Y$_x$La$_{1-x}$1111 with $x$=0.95, 0.2, and 0.
In general, $1/T_1T$ is described as $1/T_1T\propto \sum_{\bm q} |A_{\bm q}|^2 \chi''({\bm q},\omega_0)/\omega_0$, where $A_{\bm q}$ is a wave-vector (${\bm q}$)-dependent hyperfine-coupling constant and $\chi({\bm q},\omega)$ a dynamical spin susceptibility. 
Note that $1/T_1T$ is dominated by the low-energy limit of $\chi''({\bm q},\omega)/\omega$ since an NMR frequency ($\omega_0$) is as low as a radio frequency. 
In Y$_{x}$La$_{1-x}$1111 compounds, a doping level of electrons into FeAs layers is expected to be similar, because an oxygen deficiency is nearly equivalent~\cite{Yamashita_Y}. 
In most electron-doped Fe-based SCs without magnetic rare-earth ions, Knight shift that is proportional to $\chi({\bm q}=0)$ exhibits the slight decrease upon cooling\cite{GrafeNJP,Terasaki,Ning,Imai,Yamashita_Y}. 
Hence, the increase of $1/T_1T$ can be attributed to the development of low-lying AFSFs with finite ${\bm q}$ below 150 K, which is more significant in going from $x$=0 to 0.95 in Y$_{x}$La$_{1-x}$1111, as displayed in the inset.
%Since the Knight shift for La1111 as well as in other electron-doped Fe-pnictides\cite{} or similar weak $T$ dependence for Y$_{x}$La$_{1-x}$1111($x$=$0.2$)\cite{}, 
%Here, when noting that the nearly $T$ independent Knight shifts for Y$_{x}$La$_{1-x}$1111 in this $T$ range are also nearly invariant to $x$, a doping level of electrons is expected to be almost the same in these compounds~\cite{Yamashita_Y}. This is expected because an oxygen deficiency is nearly equivalent to dope electrons into FeAs layers. 
It should be noted that as the enhancement of low-lying AFSFs becomes visible upon cooling, $T_{\rm c}$ increases from 28 K to 34 K and then to 50 K at $x$=0.0, 0.2, 0.95 in Y$_{x}$La$_{1-x}$1111, respectively. 
As shown in the figure, however, this trend is not always valid when looking at the data for Al42622 ($T_{\rm c}$=27 K)\cite{Kinouchi} and BaK122 ($T_{\rm c}$=38 K)~\cite{Yashima}. Namely, $1/T_1T$s for both compounds markedly develop upon lowering temperature in association with the stronger enhancement of AFSFs than that in Y$_{0.95}$La$_{0.05}$1111, whereas $T_{\rm c}$ goes down to 38 K and 27 K for BaK122 and Al42622, respectively.  Therefore, we remark that development of low-lying AFSFs due to the FS nesting is not always a unique factor for enhancing $T_{\rm c}$, although an intimate correlation between the development of AFSFs and the enhancement of $T_{\rm c}$ is experimentally suggested in Fe-pnictides compounds such as  Ba(Fe$_{1-x}$Co$_{x}$)$_2$As$_2$(Ba122(Co))\cite{Ning}, BaFe$_2$(As$_{1-x}$P$_{x}$)$_2$(Ba122(AsP))\cite{NakaiPRB}, FeSe\cite{Imai,YShimizu}, and F-doped La1111\cite{Oka}. 

%%%%%%%%%%%%   SC state properties   %%%%%%%%%%%%%%%%

In order to shed light on SC characteristics,  $T_1(T_{\rm c})/T_1$ normalized at $T_{\rm c}$ is plotted against the normalized temperature $T/T_{\rm c}$ in Fig.~\ref{fig:T1}(b). The $T_1(T_{\rm c})/T_1$ for Y$_{0.95}$La$_{0.05}$1111 decreases steeply as $\sim T^{4\sim 5}$  without any trace of coherence peak just below $T_{\rm c}$. As shown in Fig.~\ref{fig:T1}(c), $1/T_1T$ also approaches zero at the low-$T$ limit without exhibiting a $T$-linear dependence down to $T \sim 0.1T_{\rm c}$ even under a large external field at 12 T.  These results differ from the case for the nodal SC state in $A$Fe$_2$(As$_{1-x}$P$_{x}$)$_2$($A$=Ba,Sr), which exhibits a $T_1T$=const. behavior in $T~<~0.4T_{\rm c}$, dominated by a large residual density of states (DOS) induced in nodal gaps under the large external field~\cite{NakaiPRB,Dulguun}. The $T_1$ result reveals that Y$_{0.95}$La$_{0.05}$1111 is a {\it nodeless} high-$T_{\rm c}$ superconductor with $T_{\rm c}$=50 K.

%**************  Fig. 4 **************************************
\begin{figure}[htbp]
\begin{center}
\includegraphics[width=8.5cm]{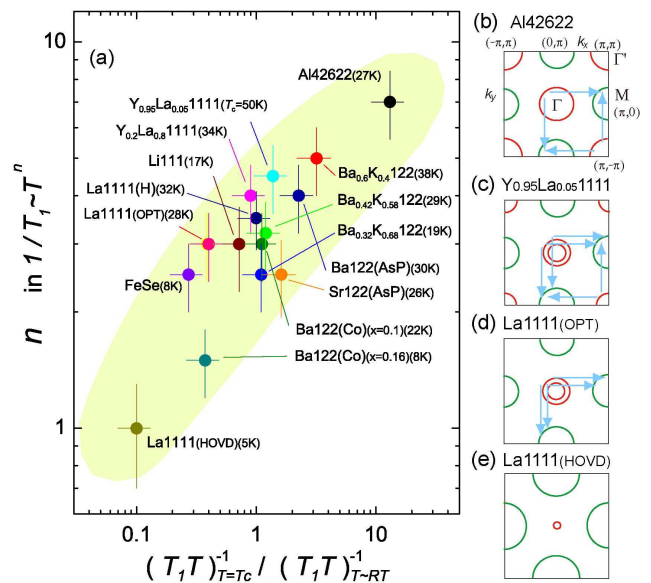}
\end{center}
\caption[]{(color online)
(a) $n$ in the formula of  $1/T_1 \sim T^n$ (SC state) is plotted against  $(T_1T)^{-1}_{T_{\rm c}}/(T_1T)^{-1}_{RT}$ (normal state), along with the data for  Ba$_{1-x}$K$_{x}$122 ($x$=0.58,0.68)\cite{Yashima_UP}, Ba122(Co)($x$=0.1\cite{Ning_JPSJ},0.16\cite{Yashima_UP}), Ba122(AsP)\cite{NakaiPRB}, Li111\cite{Li}, and FeSe\cite{Masaki_FeSe}. 
Here, this formula is assumed in the $T$ range of 0.5$T_{\rm c} < T < T_{\rm c}$. $(T_1T)^{-1}_{T_{\rm c}}$ and $(T_1T)^{-1}_{RT}$ are the values at $T$=$T_{\rm c}$ and room temperature, respectively. The plot means that as antiferromagnetic spin fluctuations are more significantly enhanced, the reduction rate in $1/T_1$ just below $T_{\rm c}$ is more remarkable without the coherence peak. Illustrations of Fermi surface topologies are suggested by the band calculations on (b) Al42622\cite{Miyake,Usui}, (c) $Ln$1111 with $T_{\rm c}$ higher than 50 K\cite{Kuroki2}, (d) La1111(OPT)\cite{Kuroki2}, and (e) La1111(HOVD). }
\label{FS}
\end{figure}
%****************************************************************************

Figure \ref{FS}(a) presents plots of the power $n$ in the formula of $1/T_1 \sim T^n$ versus  $(T_1T)^{-1}_{T_{\rm c}}/(T_1T)^{-1}_{RT}$ for various Fe-pnictide SCs.  
Here this formula is assumed in $T$ range of 0.5$T_{\rm c} < T < T_{\rm c}$ (see Fig. \ref{fig:T1}(b)). 
$(T_1T)^{-1}_{T_{\rm c}}$ and $(T_1T)^{-1}_{RT}$ are the values at $T$=$T_{\rm c}$ and room temperature, respectively.
$(T_1T)^{-1}_{T_{\rm c}}/(T_1T)^{-1}_{RT}>$1 means that AFSFs develop upon cooling, whereas $(T_1T)^{-1}_{T_{\rm c}}/(T_1T)^{-1}_{RT}<$1  points to  the band structure effect near the Fermi level(see Fig.~\ref{fig:T1}(a)). This remarkable correlation points to the fact that as AFSFs become more dominant in the normal state, the reduction rate in $1/T_1$ just below $T_{\rm c}$, i.e. $n$, increases from $n\sim 3$ for La1111 ($T_{\rm c}$=28 K) to $n\sim 7$ for Al42622 ($T_{\rm c}$=27 K)\cite{Kinouchi}, and hence the suppression of the coherence effect is more remarkable. These systematic evolutions of relaxation behaviors have allowed us to apply the multiple fully gapped $s_\pm$-wave model based on the FS nesting to various Fe-pnictides. In previous works, we have presented a simulation of the relaxation behavior below $T_{\rm c}$ with a parameter $\alpha_{\rm c}$ of the coefficient of the coherence factor within the framework of the multiple fully gapped $s_\pm$-wave model \cite{MukudaHOVD,Kinouchi}. In this simulation, $\alpha_{\rm c}=1$ is expected for sign-conserving {\it intraband} scattering and $\alpha_{\rm c}=-1$ for sign-nonconserving {\it interband} scattering. In multiband systems, this value varies in the range $-1\le \alpha_{\rm c}\le 1$, dependent on the weight of their contribution in the nuclear relaxation process. Actually, as shown by the solid lines in Fig.~\ref{fig:T1}(b), the previous experiments on  Al42622\cite{Kinouchi}, BaK122\cite{Yashima}, and La1111(HOVD)\cite{MukudaHOVD}, were reproduced by reasonable parameters with $\alpha_{\rm c}\sim -0.86$, $\sim$0, and $\sim +0.33$, respectively. Here, it is valid to assume $\alpha_{\rm c}\sim$~0 for BaK122 since other parameters of SC-gap sizes and DOS are consistent with those values deduced by ARPES\cite{Ding}. The result of Y$_{0.95}$La$_{0.05}$1111 was also reproduced by taking parameters as $2\Delta_{\rm L}/k_{\rm B}T_{\rm c}$ = 6.9$(\Delta_{\rm S}/\Delta_{\rm L}$=0.35), $N_{\rm FS^{L}}/(N_{\rm FS^{L}}+N_{\rm FS^{S}})$=0.67, smearing factor $\eta=0.05\Delta_{\rm L}$, and $\alpha_{\rm c} \sim$0. Here, SC gaps are represented as $\Delta_{\rm L}$($\Delta_{\rm S}$) and DOSs as $N_{\rm FS^{L}}$($N_{\rm FS^{S}}$) for FS with larger(smaller) gaps. 

Next, we deal with these SC and normal-state behaviors in terms  of the evolution of the Fermi surface (FS) nesting property based on band calculations reported thus far. Figures \ref{FS}(b-e) present the schematic illustration of FS topologies, which are theoretically derived on the basis of the five-orbital model for several typical compounds, such as (b) Al42622\cite{Miyake,Usui}, (c) La$_{0.05}$Y$_{0.95}$1111\cite{Kuroki2}, (d)La1111(OPT)\cite{Kuroki2}, and (e) La1111(HOVD). 
As illustrated in Figs.~\ref{FS}(d) and \ref{FS}(e), one of the hole FSs [$\Gamma^{\prime}$($\pi$,$\pi$)] sinks beneath the Fermi level ($E_{\rm F}$) as $h_{Pn}$ becomes shorter. Note that the FS nesting condition of these FSs becomes significantly worse in the heavily electron-doped La1111(HOVD) with $T_{\rm c}$=5 K, leading to the lack of AFSFs in the normal state, and as a result, $T_{\rm c}$ goes down and the coherence effect is not significantly depressed [See Fig.~\ref{FS}(e)]\cite{MukudaHOVD}. 
By contrast, when $h_{Pn}$=1.5 \AA\ in Al42622 is longer than in other Fe pnictides, the size of the hole FS around $\Gamma^{\prime}$ is larger, while one of two-hole FSs at $\Gamma$(0,0) disappears, as shown in Fig.~\ref{FS}(b)\cite{Miyake}. In this case, since the FS nesting condition is much better between hole FSs ($\Gamma$ and $\Gamma^{\prime}$) and electron FSs ($M$[(0,$\pm\pi$) and ($\pm\pi$,0)]) than in other cases\cite{Miyake,Usui}, AFSFs develop dramatically in such a manner that antiferromagnetic order could seemingly set in around 20 K. Even though the FS nesting is optimized, $T_{\rm c}$=27 K is much lower than the highest $T_{\rm c}$=55 K in Fe pnictides in association with the decrease of the FS multiplicity for Al42622, as argued below\cite{Usui}. 

Most remarkably, an important ingredient is that $Ln$1111 exhibiting $T_{\rm c}$ higher than 50 K is characterized by  three hole FSs; two of them are at $\Gamma$ and another is at $\Gamma^{\prime}$, and two electron FSs at $M$ in the unfolded FS regime, as presented in  Fig.~\ref{FS}(c). When noting that $h_{Pn}\sim$1.44 \AA\  of Y$_{0.95}$La$_{0.05}$1111\cite{Evaluation} is longer than $h_{Pn}\sim$1.35\AA\  for La$_{0.8}$Y$_{0.2}$1111($T_{\rm c}$=34 K) and  $h_{Pn}\sim$1.33\AA\ in La1111(OPT) ($T_{\rm c}$=28 K), the appearance of $\Gamma^{\prime}$ at $E_{\rm F}$ causes the FS nesting condition to be better for Y$_{0.95}$La$_{0.05}$1111 than that for La1111(OPT), resulting in the enhancement of AFSFs for the former. In this context, as the FS nesting condition becomes better, AFSFs are visible, and hence $T_{\rm c}$ increases from 28 K in La1111, to 34 K in La$_{0.8}$Y$_{0.2}$1111 up to 50 K in Y$_{0.95}$La$_{0.05}$1111.  Usui {\it et al.} have pointed out that the large FS multiplicity in $Ln$1111 in addition to the presence of AFSFs is an another crucial factor for enhancing $T_{\rm c}$ based on the spin-fluctuation mediated SC mechanism when the FeAs$_{4}$ tetrahedron is close to a regular one realized in $Ln$1111~\cite{Usui}.
In this context, the optimized electronic state for the occurrence of SC in Fe-pnictides is realized for the regular FeAs$_4$ tetrahedron where multiorbital fluctuations may play some role in the onset of SC\cite{Kontani}, since spin and orbital degrees of freedom can be intimately coupled with one another. However, it is unlikely that multiorbital fluctuations become dominant to mediate high-$T_{\rm c}$ SC in Y$_{0.95}$La$_{0.05}$1111 because they prefer an $s_{++}$ wave SC \cite{Kontani}. Furthermore, we remark that the overall $T$ dependence of $1/T_1$ in both SC and normal states is qualitatively accounted for by the multiple fully-gapped $s_\pm$-wave SC model based on the FS nesting. 

%\section{Summary}

In conclusion, the $^{75}$As-NMR $1/T_1$ measurement has revealed that Y$_{0.95}$La$_{0.05}$1111 is the nodeless high-$T_{\rm c}$ superconductor with $T_{\rm c}$=50 K taking place under the moderately enhanced AFSFs due to the FS nesting condition better than in La1111 with $T_{\rm c}$=$28$ K. We have demonstrated that these results are accounted for by the multiple fully-gapped $s_\pm$-wave model based on the the FS nesting \cite{Usui}. 
We have proposed that the reason that $T_{\rm c}\sim$50 K in Y$_{0.95}$La$_{0.05}$1111 is larger than $T_{\rm c}$=28 K in La1111 is that  the FS multiplicity is maximized with the regular FeAs tetrahedron structure, and hence the FS nesting condition is better than that of La1111, developing moderately AFSFs. In the future, to address a mechanism of high-$T_{\rm c}$ SC in Fe pnictides, it is desired to elucidate the $q$ and $\omega$ dependences of $\chi''(q,\omega)$ under the spin and orbital degrees of freedom coupled with one another for $Ln$1111 with  $T_{\rm c}$ higher than 50 K. 

%\section*{Acknowledgements}
{\footnotesize 
We thank K. Kuroki for fruitful discussion and comments. This work was supported by a Grant-in-Aid for Specially Promoted Research (20001004) and by the Global COE Program (Core Research and Engineering of Advanced Materials-Interdisciplinary Education Center for Materials Science) from the Ministry of Education, Culture, Sports, Science and Technology (MEXT), Japan.
}

%::::::::::::::::bibliography::::::::::::::::::::::::::::::::::::::::::::::::
%:::::::::::::::::::::::::::::::::::::::::::::::::::::::


\begin{thebibliography}{99}
\bibitem{Evaluation} The $h_{Pn}$ for Y$_{x}$La$_{1-x}$1111 was deduced by an assumption that the bond length of Fe-As is invariant among Y$_{x}$La$_{1-x}$1111\cite{C.H.Lee}.
\end{thebibliography}
\end{document}